\documentstyle[epsf]{mn}

\def\astrobj#1{#1}
\def\url#1{{\ttfamily\def\/{/\discretionary{}{}{}}#1}}
\def\harvarditem#1#2#3{\bibitem[#2]{#3}}
\def\bibcode#1{}
\def\zijlstra{Zijl\-stra}

\def\ulapskip#1#2{\vbox to0pt{\parindent=0mm\vss\hskip#1 #2}}
\def\dot#1{\leavevmode\setbox0\hbox{h}\dimen0\ht0\advance\dimen0-0.75ex
   \rlap{\raise1.40\dimen0\ulapskip{0.39em}{\Large$\cdot$}}#1}
\newcommand{\dm}{\relax\ifmmode{\dot{M}}\else{$\dot{M}$}\fi}
\newcommand{\teff}{\relax\ifmmode{T_{\rm eff}}\else{$T_{\rm eff}$}\fi}
\newcommand{\tstar}{\relax\ifmmode{T_{\ast}}\else{$T_{\ast}$}\fi}

\newcommand{\pta}{\relax\ifmmode{P_{\rm a}}\else{$P_{\rm a}$}\fi}
\newcommand{\ptb}{\relax\ifmmode{P_{\rm b}}\else{$P_{\rm b}$}\fi}
\newcommand{\md}{\mbox{\rm d}}

\newcommand{\p}{{\mathchoice
  {\setbox0=\hbox{$\displaystyle'$}\kern0.25\wd0\lower0.28\ht0\box0}
  {\setbox0=\hbox{$\textstyle'$}\kern0.25\wd0\lower0.28\ht0\box0}
  {\setbox0=\hbox{$\scriptstyle'$}\lower0.28\ht0\box0}
  {\setbox0=\hbox{$\scriptscriptstyle'$}\lower0.28\ht0\box0}}}
\newcommand{\pp}{\relax\ifmmode{^\prime}\else{$^\prime$}\fi}
\newcommand{\binom}[2]{\left(\!\!\begin{array}{c} #1 \\ #2 \end{array}\!\!\right)}

\newcommand{\rs}{r_{\rm s}}
\newcommand{\ri}{r_{\rm i}}
\newcommand{\rii}{\rs}
\newcommand{\Phd}{\Phi_{\rm d}}
\newcommand{\Phb}{\Phi_{\rm b}}
\newcommand{\dia}{\Theta_{\rm d}}
\newcommand{\xt}[1]{\mbox{$\times 10^{#1}$}}
\newcommand{\kms}{\relax\ifmmode{\rm km\,s^{-1}}\else{km\,s$^{-1}$}\fi}

\newcommand{\mic}{\relax\ifmmode{\mu{\rm m}}\else{$\mu$m}\fi}
\newcommand{\zm}{\relax\ifmmode{\rm M_\odot}\else{M$_\odot$}\fi}
\newcommand{\zl}{\relax\ifmmode{\rm L_\odot}\else{L$_\odot$}\fi}
\newcommand{\zs}{\relax\ifmmode{\rm R_\odot}\else{R$_\odot$}\fi}
\newcommand{\zmy}{\relax\ifmmode{\rm M_\odot\,yr^{-1}}\else{M$_\odot$\,yr$^{-1}$}\fi}
\newcommand{\ky}{\relax\ifmmode{\rm K\,yr^{-1}}\else{K\,yr$^{-1}$}\fi}
\newcommand{\wsm}{\relax\ifmmode{\rm W\,m^{-2}}\else{W\,m$^{-2}$}\fi}
\newcommand{\escs}{\relax\ifmmode{\rm erg\,cm^{-2}\,s^{-1}}\else{erg\,cm$^{-2}$\,s$^{-1}$}\fi}
\newcommand{\rscs}{\relax\ifmmode{\rm Ryd\,cm^{-2}\,s^{-1}}\else{Ryd\,cm$^{-2}$\,s$^{-1}$}\fi}
\newcommand{\fhb}{\relax\ifmmode{F\rm (H\beta)}\else{$F$(H$\beta$)}\fi}
\newcommand{\lhb}{\relax\ifmmode{L\rm (H\beta)}\else{$L$(H$\beta$)}\fi}
\newcommand{\jb}{{\rm\sl B}}
\newcommand{\jv}{{\rm\sl V}}

\newcommand{\av}{\relax\ifmmode{A_{\rm V}}\else{$A_{\rm V}$}\fi}
\newcommand{\ebv}{\relax\ifmmode{E(\jb-\jv)}\else{$E(\jb-\jv)$}\fi}

\newcommand{\w}{\relax\ifmmode{\lambda}\else{$\lambda$}\fi}
\newcommand{\lya}{\relax\ifmmode{\rm Ly\alpha}\else{\rm Ly$\alpha$}\fi}
\newcommand{\ha}{\relax\ifmmode{\rm H\alpha}\else{\rm H$\alpha$}\fi}
\newcommand{\hb}{\relax\ifmmode{\rm H\beta}\else{\rm H$\beta$}\fi}
\newcommand{\hg}{\relax\ifmmode{\rm H\gamma}\else{\rm H$\gamma$}\fi}
\newcommand{\hd}{\relax\ifmmode{\rm H\delta}\else{\rm H$\delta$}\fi}
\newcommand{\he}{\relax\ifmmode{\rm H\epsilon}\else{\rm H$\epsilon$}\fi}
\newcommand{\bra}{\relax\ifmmode{\rm Br\alpha}\else{\rm Br$\alpha$}\fi}
\newcommand{\brg}{\relax\ifmmode{\rm Br\gamma}\else{\rm Br$\gamma$}\fi}
\newcommand{\pfa}{\relax\ifmmode{\rm Pf\alpha}\else{\rm Pf$\alpha$}\fi}

\newcommand{\fb}[2]{[#1\,{\sc #2}]}

\newcommand{\al}[2]{#1\,{\sc #2}}
\newcommand{\odd}{\rlap{\rm o}}
\newcommand{\od}[3]{\smash{\relax\ifmmode{^{#1}{\rm #2}^{\odd}_{#3}}\else{$^{#1}$#2$^{\odd}_{#3}$}\fi}}
\newcommand{\ev}[3]{\smash{\relax\ifmmode{^{#1}{\rm #2}_{#3}}\else{$^{#1}$#2$_{#3}$}\fi}}

\newcommand{\fwhm}{{\sc fwhm}}

\newcommand{\str}{Str\"omgren}

\newcommand{\eq}{Eq.}

\newcommand{\req}[1]{\eq~(\ref{#1})}

\newcommand{\reqb}[1]{\eq~\ref{#1}}

\newcommand{\sct}{Section}
\newcommand{\scts}{Sections}
\newcommand{\fig}{Fig.}

\newcommand{\tbl}{Table}

\newcommand{\cld}{{\sc cloudy}}

\newcommand{\clean}{{\sc clean}}

\newcommand{\sctr}[1]{\multicolumn{1}{c}{#1}}
\newcommand{\ctr}[3]{\multicolumn{#1}{c}{#2\hspace*{#3mm}}}

\newcommand{\m}{\rlap{$^{\dagger}$}}

\newcommand{\f}{\phantom{0}}

\newcommand{\x}[1]{\hspace*{#1mm}}
\newdimen\mbtindent
\def\mbtlist#1{\setbox0=\hbox{#1\,--\,}\mbtindent=\wd0\vspace{\baselineskip}
\bgroup\hsize=\columnwidth\leftskip=\mbtindent%
\rightskip=0pt\parindent=-\mbtindent\lineskiplimit=-10pt\parskip=0pt}
\def\endmbtlist{\egroup}
\newcommand{\mbt}[2]{\leavevmode\hbox to\mbtindent{#1\hfil--\,}#2\par}

\def\nt#1{\vtop{\footnotesize\hsize=\columnwidth\leavevmode#1\hspace*{\fill}}}

\def\form{3}
\def\alter{4}
\def\secmom{5}

\begin{document}

\title[Measuring angular diameters of extended sources]
{Measuring angular diameters of extended sources}
\author[P.A.M. van Hoof]{P.A.M. van Hoof\thanks{Currently staying at CITA,
University of Toronto, Canada.}\\
Kapteyn Astronomical Institute, P.O. Box 800, 9700 AV Groningen,
The Netherlands\\
University of Kentucky, Department of Physics and Astronomy,
177 CP Building, Lexington, KY 40506--0055, USA}
\date{received, accepted}
\maketitle

\begin{abstract}
When measuring diameters of partially resolved sources like planetary nebulae,
\al{H}{ii} regions or galaxies, often a technique called gaussian
deconvolution is used. This technique yields a gaussian diameter which
subsequently has to be multiplied with a conversion factor to obtain the true
angular diameter of the source. This conversion factor is a function of the
\fwhm\ of the beam or point spread function and
also depends on the intrinsic surface brightness
distribution of the source.

In this paper conversion factors are presented for a number of simple
geometries: a circular constant surface brightness disk and a spherical
constant emissivity shell, using a range of values for the inner radius. Also
more realistic geometries are studied, based on a spherically symmetric
photo-ionization model of a planetary nebula. This enables a study of optical
depth effects, a comparison between images in various emission lines and the
use of power law density distributions. It is found that the conversion factor
depends quite critically on the intrinsic surface brightness distribution,
which is usually unknown. The uncertainty is particularly large if extended
regions of low surface brightness are present in the nebula.
In such cases the use of gaussian or second moment deconvolution is not
recommended.

As an alternative, a new algorithm is presented which allows the determination
of the intrinsic \fwhm\ of the source using only the observed surface
brightness distribution and the \fwhm\ of the beam. Hence no assumptions
concerning the intrinsic surface brightness distribution are needed. Tests
show that this implicit deconvolution method works well in realistic
conditions, even when the signal-to-noise is low, provided that the beam size
is less than roughly $2/3$ of the observed \fwhm\ and the beam profile can be
approximated by a gaussian. A code implementing this algorithm is
available.
\end{abstract}

\begin{keywords}
Methods: data analysis --- ISM: general
\end{keywords}

\section{Introduction}

The accurate measurement of angular diameters is a long standing problem. It
is pertinent to the study of planetary nebulae, \al{H}{ii} regions, galaxies
and other extended sources. Nevertheless, only few papers dedicated to this
problem can be found in the literature, e.g.\ Mezger \& Henderson
\cite{c3:mezger}, Panagia \& Walmsley \cite{c3:pana}, Bedding \& \zijlstra\
\cite{c3:bed:zijl}, Schneider \& Buckley \cite{c3:schneider}, and Wellman et
al. \cite{c3:wellman}. This paper will be written in the context of planetary
nebula research. However, most results will also be valid in a more general
context.

Several methods are in general use to determine angular diameters. For nebulae
with a well-defined outer radius (i.e.\ with a steep drop-off to zero surface
brightness at a certain radius) it is easy to measure directly the radius
where a prescribed value of the surface brightness is reached. This prescribed
value often is a certain fraction of the peak surface brightness (usually
10~\%). This method will be called direct measurement. It is in general use
for observations of well-resolved sources and will not be studied in this
paper. This method works, provided that observations of sufficient resolution
and quality are available. This way no assumptions have to be made about the
intrinsic surface brightness distribution of the source and this explains the
popularity of this method. It should be pointed out that in {\em all} other cases
(i.e.\ when the source is not well resolved or when it does not have a
well-defined outer radius) assumptions have to be made about the intrinsic
surface brightness distribution in order to interpret the results.
In the remainder of the paper we will also refer to the surface brightness
distribution as surface brightness profile or simply profile.

For observations where the source is only partially resolved, one has to
resort to different methods.
One method is based on the Full Width at Half Maximum
(\fwhm) of a two-dimensional gaussian fitted to the observed surface
brightness distribution in a least-squares sense. This method is usually
called gaussian deconvolution and will be explained in more detail below.
Another method that is being used is basically identical to the first, except
that it determines the \fwhm\ using the second moment of the surface
brightness distribution instead of a gaussian fit. To discriminate it from the
first method, it will be called second moment deconvolution. The choice for
either method depends mainly on the preference of the observer. Both methods
have the disadvantage that they yield a result that has no direct physical
meaning. Hence, a conversion factor is needed to translate the result into
something meaningful. In nebular research this usually is the \str\ radius.
This conversion factor depends on the method being used, the intrinsic surface
brightness distribution of the source and the size of the beam.

Another problem is that not all nebulae have a well-defined outer radius.
Often, when deeper images are made, more emission is detected at lower surface
brightness levels. Such nebulae will be referred to as having a soft boundary.
For such nebulae the 10~\% radius (or a radius at any other percentage level)
does not have a direct physical meaning and does not represent the nebular
size very well. The radius becomes increasingly larger when deeper images are
taken. The \str\ radius can usually not be observed directly either. Hence the
size of such nebulae cannot be represented in a meaningful way by a single
number. In \sct~\ref{non:constant} it will be shown that also the application
of gaussian or second moment deconvolution to such nebulae leads to large
uncertainties and cannot be recommended.

The major disadvantage of gaussian and second moment deconvolution is that
assumptions have to be made about the shape (but not the size) of the
intrinsic surface brightness distribution of the source. It will be shown in
this paper that this choice is quite critical. However, when only low
resolution observations are available one can make no more than an educated
guess about this distribution. As an alternative, a new algorithm is presented
which allows the determination of the intrinsic \fwhm\ of the source using
only the observed surface brightness distribution and the \fwhm\ of the beam.
Hence the major advantage of this method is that it yields a deconvolved
diameter without necessitating assumptions concerning the intrinsic surface
brightness distribution. This process is {\em not} an image reconstruction
algorithm and therefore requires far less computational overhead. It should be
pointed out that it can only give the \fwhm\ diameter and not the \str\
diameter. For the latter conversion, assumptions concerning the shape of the
nebula will always be necessary.

This article will have the following structure: in \sct~\ref{comput} some
basic assumptions and definitions will be given. Also, the methods used to
calculate the conversion factors will be discussed. In \sct~\ref{simple}
conversion factors will be given for various simple geometries: a circular
constant surface brightness disk and a spherical constant emissivity shell,
using a range of values for the inner radius. In \sct~\ref{previous} these
results will be compared to previous studies found in the literature and a
discussion will be given. Next, the conversion factors will be studied using
more realistic geometries based on a photo-ionization model of a planetary
nebula. In \sct~\ref{op:depth} the influence of optical depth effects on the
observed surface brightness distribution
and on the conversion factor will be studied. In \sct~\ref{hydrogen} images
constructed in several optical emission lines will be compared and the influence on
the conversion factor will be discussed. In \sct~\ref{non:constant} the effect
of non-constant density laws on the conversion factor will be studied. These
density laws allow a discussion of the appropriateness of gaussian or second
moment deconvolution for nebulae with a soft boundary. In
\sct~\ref{new:method} a new method will be presented which allows the
determination of the intrinsic \fwhm\ of a profile, using only the observed
profile and the beam size. Finally, in \sct~\ref{concl:iv} the main
conclusions will be presented. The theory used to calculate the conversion
factors has been presented in van Hoof \cite{vh1} (hereafter Paper~I). This
paper is available through the e-print archive at \url{http://xxx.lanl.gov}
under number astro-ph/9906051.

\section{Definitions and computational methods}
\label{comput}

The methods discussed in this article can be applied to observations at any
wavelength. More in particular, they are valid for optical, infrared and radio
observations. The resolution of these observations is usually characterized by
the size of the beam profile for radio data and by the size of the point
spread function for optical or infrared data. Throughout the paper the term
`beam' will be used and it will be implicitly understood that it can also mean
`point spread function' where appropriate. It will be assumed that the beam
can be approximated by a gaussian. This is a reasonable assumption, both for
radio and for optical observations. First, in the reduction of radio
observations, the (possibly complicated) antenna pattern of the telescope is
replaced by a perfect gaussian of the same resolution in the \clean\
procedure. Second, for optical observations the point spread function is
normally determined by the seeing which can be approximated by a gaussian.
This approximation is however only valid for the core region of the point
spread function, further out it is better represented by an inverse square law
(e.g., King 1971). This implies that care should be taken when interpreting
low level emissions surrounding barely resolved nebulae; accurate knowledge of
the point spread function is required in such cases (Falomo 1996). In this
paper the intrinsic surface brightness profile will be defined as the surface
brightness distribution that would be observed with infinite resolving power.
For simplicity it will be assumed throughout this paper that both the surface
brightness distribution of the nebula and the beam are circularly symmetric.
This is a rather severe restriction; nebulae rarely are circular and also, for
radio observations, the beam usually is elliptical. However, this simplified
case already yields interesting results which can be applied to actual data.
Since both the intrinsic profile of the nebula and the beam profile are
assumed to be circularly symmetric, they can
be represented
as one-dimensional functions
measuring the profile radially outwards from the centre.

As was already remarked, a conversion factor is needed to translate the \fwhm\
diameter yielded by gaussian or second moment deconvolution into a \str\
diameter. In this paper the \str\ radius of the nebula will be denoted by
$\rs$,
and the true diameter by $\dia = 2\rs$. The \fwhm\ of the observed
nebular image will be denoted by $\Phi$ and the \fwhm\ of the beam by $\Phb$.
Throughout the paper the deconvolved \fwhm\ diameter $\Phd$ will be used,
which is defined by
\begin{equation}
  \Phd = \sqrt{\Phi^2 - \Phb^2}.
\label{phd:def}
\end{equation}
This quantity is also commonly called the gaussian diameter. The conversion
factor to obtain the true angular diameter from the deconvolved \fwhm\ can
be defined as
\begin{equation}
  \dia = \gamma \Phd \x{3} \Rightarrow \x{3} \gamma = \dia/\Phd = 2\rs/\Phd.
\label{convfac}
\end{equation}
The deconvolved \fwhm\ should {\em not} be confused with the \fwhm\ of the
deconvolved profile, which in general will not be equal. The latter will be
called the intrinsic \fwhm. The conversion factor $\gamma$ is a function of
the resolution of the observation, or to be more precise, of the ratio of the
source diameter and the beam size. Hence an independent parameter $\beta$ is
chosen, which is defined as
\begin{equation}
  \beta = \Phd/\Phb.
\label{beta:def}
\end{equation}
In the following sections more details will be given of the techniques that
have been used to calculate the conversion factors, both for gaussian and
second moment deconvolution.

\subsection{Gaussian deconvolution}
\label{comput:gauss}

First the technique that has been used to calculate conversion factors for
gaussian deconvolution will be discussed. This technique is based
on an implicit equation from which the value of $\gamma(\beta)$ can be solved
for arbitrary $\beta$. The derivation of this expression has been presented in
\sct~\form\ of Paper~I. To use this technique, first the radial moments $c_n$ of
the assumed intrinsic surface brightness profile $f(r)$ have to be computed.
They are defined as:
\[   c_n = 2\pi\!\int_0^\infty f(r) r^{n+1} \md r, \]
and are computed using either an analytic expression (if available) or a
numerical integration scheme. Next these radial moments and also the value for
$\beta$ are substituted in \req{gen:eqn}.
\[ \sum_{n=0}^\infty \frac{c_{2n} \gamma^{2n}(\beta) \ln^n2}{n!}
   \sum_{k=0}^n\,(-1)^{n-k}\binom{n}{k} \times
\]
\begin{equation}
\label{gen:eqn}
   \phantom{\sum_{n=0}^\infty \frac{c_{2n} \gamma^{2n}(\beta)\ln^n}{n!}}
   \times (\beta^2-2k) \, \beta^{2n-2} \left( \frac{\beta^2+1}{\beta^2+2}
   \right)^k = 0.
\end{equation}
The value for the conversion factor $\gamma$ is solved iteratively using a
Newton-Raphson scheme. If this procedure is repeated for a range of values for
$\beta$, the behaviour of the conversion factor $\gamma(\beta)$ as a function
of $\beta$ can be found.

In order to represent the results efficiently, a simple analytic function will
be fitted to the conversion factors. Tests have shown that $\gamma(\beta)$ can
be approximated extremely well by the following function
\begin{equation}
  \gamma_f(\beta) = \frac{a_{1}}{1 + a_{2}\beta^{2}} + a_{3},
\label{fitfunc}
\end{equation}
which will be used throughout this paper. The fit is determined by minimizing
the reduced $\chi^2$ which is defined as
\begin{equation}
  \chi^2 = \frac{1}{N}\sum_{n=1}^N\,\left[\,\gamma(\beta_n)-
  \gamma_f(\beta_n)\,\right]^2.
\label{chi2:def}
\end{equation}
Here $N$ is the total number of points at which the conversion factor has been
evaluated (usually 251, with $\beta_n = 0\,(0.02)\,5$)\footnote{This notation
indicates that $\beta_n$ runs from 0 to 5 with a step size of 0.02.}.

\subsection{Second moment deconvolution}
\label{comput:secmom}

Second moments are widely used to calculate the \fwhm\ of an arbitrary
profile. In general however, the result of this method will not be identical
to the \fwhm\ derived from a gaussian fit. Hence, also the value for the
conversion factor will be different. To distinguish the results of the two
methods, a subscript $2$ will be used on all quantities derived with the
second moments method. In Paper~I, \sct~\secmom\ it has been proven that when
second moment deconvolution is used, the conversion factor is constant (i.e.\
{\em independent} of beam size). Furthermore, this constant value is equal to
the value of the conversion factor for gaussian deconvolution in the limit for
infinitely large beams. In other words, the conversion factor for second
moment deconvolution is given by
\[ \gamma_2(\beta) = \gamma(0) \x{3} {\rm for\ all\ }
   \beta\in[\,0,\infty). \]
Due to this relation it is not strictly necessary to give separate results.
The value for the conversion factor for second moment deconvolution can always
be calculated using the following expression
\begin{equation}
  \gamma_2 \approx \gamma_f(0) = a_1 + a_3.
\label{gam2use}
\end{equation}
Given the quality of the fitting function $\gamma_f$, this yields results with
more than sufficient accuracy (a few times 10$^{-3}$ down to a few times
10$^{-4}$).

\section{The conversion factor for simple geometries}
\label{simple}

The methods which have been discussed in the previous section are
applied to images which are only partly resolved. Hence they contain little
direct information on the intrinsic surface brightness profile. If no other
information is available, the intrinsic surface brightness profile has to be
assumed in order to calculate the conversion factor. In such cases, the choice
is usually a very simple geometry. In this section, conversion factors will
be determined for the geometries that where
presented in Bedding \& \zijlstra\ \cite{c3:bed:zijl}.
These are the constant surface brightness disk (the limiting case of a
spherically symmetric nebula which is completely optically thick, or
alternatively a nebula with cylindrical symmetry viewed along the axis), and
the constant volume emissivity shell and sphere (the limiting case of a
spherically symmetric nebula with zero optical thickness). These cases will be
treated in more detail here, and also shells with an arbitrary inner radius
will be treated. The sphere can be viewed as the limiting case of a shell with
zero inner radius.

\begin{table}
\caption
{The parameters for calculating conversion factors. Results are given for a
constant surface brightness disk and for a constant emissivity shell with
various ratios of the inner to outer radius (as indicated in column~1). The
conversion factors for gaussian deconvolution can be calculated using
\req{fitfunc}, the conversion factors for second moment deconvolution are
given separately in the last column.}
\label{simpletab}
\begin{tabular}{llllll}
\hline
case & \sctr{$a_{1}$} & \sctr{$a_{2}$} & \sctr{$a_{3}$} & \sctr{$\chi^{2}$} & \sctr{$\gamma_2$} \\
\hline
disk      &   0.3512 &   0.7874 &   1.3469 & 5.4($-$8)\m& 1.6986 \\
\\
shell 0.0 &   0.3358 &   0.7907 &   1.5629 &  4.1($-$8) & 1.8991 \\
shell 0.1 &   0.3368 &   0.7896 &   1.5609 &  4.3($-$8) & 1.8982 \\
shell 0.2 &   0.3431 &   0.7835 &   1.5482 &  5.1($-$8) & 1.8918 \\
shell 0.3 &   0.3563 &   0.7745 &   1.5187 &  6.3($-$8) & 1.8756 \\
shell 0.4 &   0.3729 &   0.7694 &   1.4733 &  6.8($-$8) & 1.8468 \\
shell 0.5 &   0.3875 &   0.7715 &   1.4168 &  6.5($-$8) & 1.8049 \\
shell 0.6 &   0.3959 &   0.7786 &   1.3546 &  6.1($-$8) & 1.7510 \\
shell 0.7 &   0.3963 &   0.7871 &   1.2909 &  5.9($-$8) & 1.6877 \\
shell 0.8 &   0.3893 &   0.7940 &   1.2281 &  5.7($-$8) & 1.6180 \\
shell 0.9 &   0.3766 &   0.7981 &   1.1678 &  5.4($-$8) & 1.5449 \\
limit 1.0 &   0.3600 &   0.7994 &   1.1106 &  5.0($-$8) & 1.4711 \\
\hline\multicolumn{5}{l}
{\m\x{4}5.4($-$8) stands for 5.4\xt{-8}.}\\
\end{tabular}
\end{table}

\begin{table}
\caption{The conversion factor and the \fwhm\ for the unconvolved disk and
shell profiles.}
\label{tab:inf}
\begin{tabular}{lll}
\hline
case      &\sctr{$\gamma(\infty)$}&  \sctr{$\Phi/\rs$}  \\
\hline
disk      & 1.346\,346 & 1.485\,502 \\
\\
shell 0.0 & 1.562\,397 & 1.280\,084 \\
shell 0.1 & 1.560\,415 & 1.281\,710 \\
shell 0.2 & 1.547\,616 & 1.292\,310 \\
shell 0.3 & 1.518\,086 & 1.317\,449 \\
shell 0.4 & 1.472\,623 & 1.358\,121 \\
shell 0.5 & 1.416\,124 & 1.412\,306 \\
shell 0.6 & 1.353\,972 & 1.477\,135 \\
shell 0.7 & 1.290\,243 & 1.550\,095 \\
shell 0.8 & 1.227\,504 & 1.629\,323 \\
shell 0.9 & 1.167\,188 & 1.713\,520 \\
limit 1.0 & 1.110\,004 & 1.801\,795 \\
\hline
\end{tabular}
\end{table}

\subsection{The constant surface brightness disk}
\label{csbd}

\begin{figure}
\begin{center}
\mbox{\epsfxsize=0.45\textwidth\epsfbox[78 327 497 720]{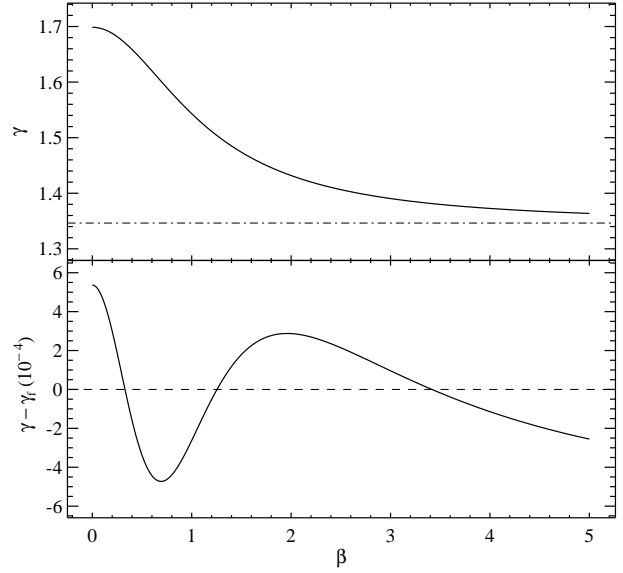}}
\caption
{The conversion factors for a constant surface brightness disk (upper panel)
and the residuals of the fit to these conversion factors (lower panel). The
dash-dotted line in the upper panel indicates the value for $\gamma(\infty)$.}
\label{diskfig}
\end{center}
\end{figure}

The conversion factors for gaussian deconvolution are shown in
\fig~\ref{diskfig} and the parameters for the fit are given in
\tbl~\ref{simpletab}. The residuals of the fit are also shown in
\fig~\ref{diskfig}. Additionally, the conversion factor for second moment
deconvolution is given in \tbl~\ref{simpletab}. In \tbl~\ref{tab:inf} the
conversion factor and the \fwhm\ for the unconvolved profile are given. These
numbers are intended as benchmarks and can be useful for testing gaussian fit
algorithms. The values were calculated using \req{gen:inf}, which can be solved
using a Newton-Raphson scheme.
\begin{equation}
   \sum_{n=0}^\infty (-1)^n\,(2n+1)\,\frac{c_{2n}\,\gamma^{2n}\,\ln^n2}
   {n!} = 0; \x{3} \frac{\Phi}{\rs} = \frac{2}{\gamma}.
\label{gen:inf}
\end{equation}
This formula can derived from \req{gen:eqn} by taking the limit
$\beta \rightarrow \infty$. It constitutes a new way of measuring the \fwhm\
of an observed profile and is discussed in more detail in Paper~I.
The values given in \tbl~\ref{tab:inf} are accurate in all decimal places.

\subsection{The constant emissivity shell}

\begin{figure}
\begin{center}
\mbox{\epsfxsize=0.45\textwidth\epsfbox[105 329  467 688]{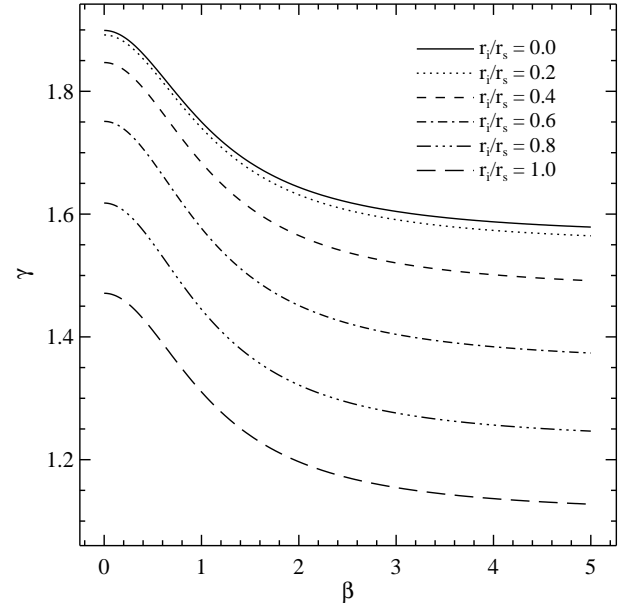}}
\caption
{The conversion factor for constant emissivity shells, for various ratios of
the inner to outer radius ($\ri/\rii$).}
\label{shellplot}
\end{center}
\end{figure}

Conversion factors were also computed for the case of a constant emissivity
shell with a range of values for the ratio of the inner to outer radius:
$\ri/\rii$ = 0.0 (0.1) 1.0. The results are shown in \fig~\ref{shellplot} and
the fit parameters are given in \tbl~\ref{simpletab}. Again the quality of the
fits is very good. It can be seen that the shape of the curve doesn't change
very much as a function $\ri/\rii$, but that the height of the curve does.
This means that the correct value for the conversion factor does depend quite
critically on the assumed value for the inner radius. The conversion factors
for second moment deconvolution are also given in \tbl~\ref{simpletab}. The
conversion factor and the \fwhm\ for the unconvolved profiles are given in
\tbl~\ref{tab:inf}. See also the remarks in \sct~\ref{csbd}.

\section{Comparison of the results}
\label{previous}

In this section the results from \sct~\ref{simple} will be compared with other
results. First, the conversion factors will be compared to the results from a
more straightforward approach. This is done to verify
the correctness of the procedure used in this paper. This comparison will be
limited to the geometries that were already studied in the literature: the
constant surface brightness disk, the constant emissivity sphere and the
constant emissivity shell with $\ri/\rii$ = 0.8. Next, the results from this
paper will be compared to published data.

\subsection{Comparison with a different technique}

In order to verify the procedure described in \sct~\ref{comput:gauss} for
calculating conversion factors for gaussian deconvolution, a more
straightforward approach was used to check the results. This
technique essentially mimics the procedure used in real observations: a given
intrinsic surface brightness profile is convolved with a gaussian of
prescribed width and subsequently gaussian deconvolution is applied using
a standard gaussian fit routine.
In order to distinguish the two techniques, the procedure
described in \sct~\ref{comput:gauss} will be called method~A, and the
procedure described here will be called method~B. The reason that method~A was
adopted throughout the paper is that it is by far the
fastest and most accurate method.

For the three geometries that have been mentioned above, the results of
method~A were found to be in excellent agreement with the results of method~B.
For the disk case it was found that $\chi^2$ = 4.6\xt{-7}, for the shell case
$\chi^2$ = 7.4\xt{-8} and for the sphere case $\chi^2$ = 3.0\xt{-8}
(see \reqb{chi2:def}). This proves the correctness of method~A. The comparison
is also visualized in \fig~\ref{compfig}.

Since the calculation of the conversion factors for second moment
deconvolution is closely linked to the calculation of the conversion factors
for gaussian deconvolution (see \sct~\ref{comput:secmom}), this also proves
the correctness of our results for the second moment method.

\subsection{Comparison with published results}

\begin{figure}
\begin{center}
\mbox{\epsfxsize=0.45\textwidth\epsfbox[163 339  423 720]{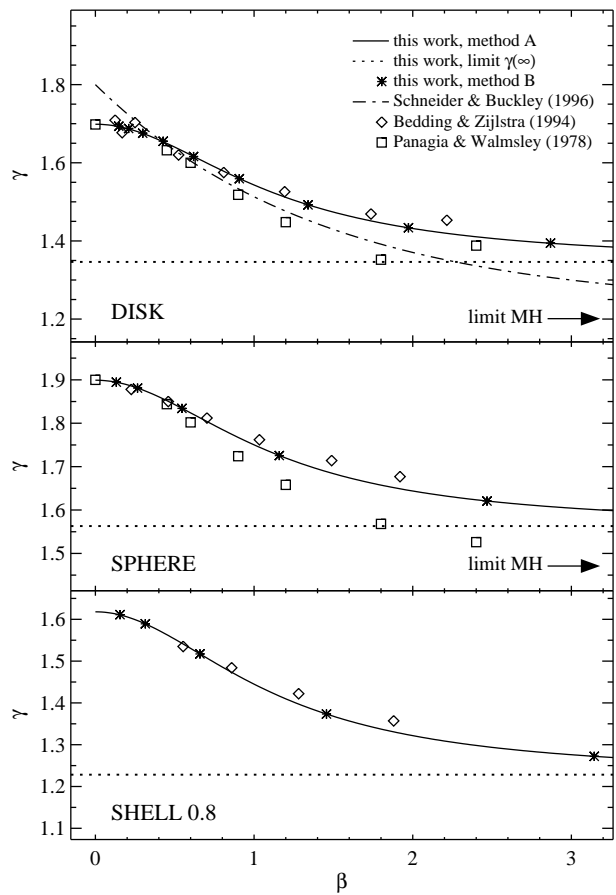}}
\caption{Comparison of this work with previous results. Methods A and B are
explained in the text. The arrows `limit MH' indicate the value for
$\gamma(\infty)$ obtained by Mezger \& Henderson \protect\cite{c3:mezger}.}
\label{compfig}
\end{center}
\end{figure}

Four papers have been previously published which were (at least in part)
dedicated to calculating conversion factors
for circularly symmetric profiles. These are Mezger \& Henderson
\cite{c3:mezger} (hereafter MH), Panagia \& Walmsley \cite{c3:pana} (hereafter
PW), Bedding \& \zijlstra\ \cite{c3:bed:zijl} (hereafter BZ) and Schneider \&
Buckley \cite{c3:schneider} (hereafter SB). A comparison of all the results is
shown in \fig~\ref{compfig}.
It should be noted that in
all these papers the gaussian deconvolution method was used. Second moment
deconvolution has never been studied before in the literature and hence no
comparison can be given for the results of this method.

Mezger \& Henderson \cite{c3:mezger} assumed that conversion factors are
independent of the beam size. They derived the conversion factors by comparing
the diameter of an unconvolved disk or sphere to a gaussian with the same peak
surface brightness and total flux. Hence one could say that they assumed
$\beta=\infty$. However their method is clearly not appropriate, since fitting
a gaussian to a given profile does not conserve the flux nor does it conserve
the peak surface brightness. It can be seen that their results are
substantially lower than the results from this study.

Panagia \& Walmsley \cite{c3:pana} re-examined the conversion factors for both
geometries. They concluded that the conversion factors depend on the beam size
and that adopting the results of MH will generally lead to an underestimation
of the nebular diameter. The value of the conversion factor for $\beta=0$ was
calculated using an analytic expression. The method they used to calculate the
other points is not clearly described.
It can be seen that for $\beta=0$ their results coincide with the
results of this paper. However, for larger $\beta$ an increasing discrepancy
between the results becomes apparent. In the disk case they even find a
minimum in $\gamma(\beta)$ which is not reproduced in this work. These
discrepancies will be discussed together with the results of SB.

\begin{figure}
\begin{center}
\mbox{\epsfxsize=0.45\textwidth\epsfbox[24 318  534 669]{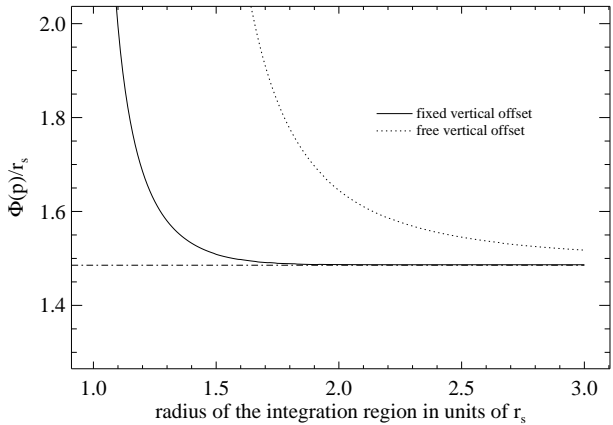}}
\caption{The \fwhm\ resulting from a gaussian fit to a constant surface
brightness disk as a function of the radius of the fitting region. The
dash-dotted line indicates the value of the \fwhm\ for an infinitely large
fitting region, i.e.\ the correct value given in \tbl~\protect\ref{tab:inf}.}
\label{warnfig}
\end{center}
\end{figure}

Bedding \& \zijlstra\ \cite{c3:bed:zijl} re-examined the conversion factors for
the disk and the sphere case and also added the case of a shell geometry. The
method they used was essentially identical to method~B. Dr.\ \zijlstra\ kindly
provided me with a table from which the values of $\beta$ and $\gamma$ could
be determined. These results are clearly the most accurate of all literature
data. It seems BZ systematically make a slight overestimation of the
conversion factor for larger values of $\beta$. An explanation for this is not
apparent to the author.

Schneider \& Buckley \cite{c3:schneider} studied only the disk geometry. They
were the first to use an analytic fit to the conversion factor as a function
of $\beta$. The method they used was essentially identical to method~B. In
their results it can be seen that the conversion factor is progressively
underestimated for larger $\beta$. The most likely explanation for this result
(and also for the results of PW) is as follows. The gaussian fit to the
(convolved) surface brightness profile is determined by minimizing the
quadratic residuals, which is defined as an integral over an infinite region.
In a numerical code this integral is replaced by an integral over a finite
region, where the upper limit of this region should be large enough not to
influence the result. If the upper limit is chosen too low, the fit will not
be `punished' for the tails of the gaussian outside the integration region. In
the case of a disk geometry this will lead to an overestimation of the \fwhm\
as is shown in \fig~\ref{warnfig} (solid curve).
In view of this result, it is recommended
to use a diameter for the fitting region which is at least 3 times the \fwhm\
diameter of the fit, irrespective of the fact if the nebula has a non-zero
surface brightness in the whole of this region or not. For nebulae with
extended faint emission the integration region should encompass the whole of
the nebula of course. In the case where the surface brightness profile is a
perfect gaussian (which is the case in the limit for infinitely large beam
sizes, $\beta=0$), the mentioned effect does not exist. The residuals will be
zero everywhere and the upper limit of the integration region is irrelevant.
The bigger the discrepancy between the actual profile and a perfect gaussian
is (i.e.\ the larger $\beta$ is), the stronger the fit will be affected by the
effect. This is exactly what can be seen in the results of PW and SB. The
local minimum that was found by PW for the disk case may have been caused by
the fact that they used a larger integration region for the last point,
although this is not clear from their paper.

Another point of caution is the following. In order to obtain an accurate
value for the \fwhm\ of an observed source, it is essential that the global
background in the image is subtracted first so that it can assumed to be zero
in the fitting procedure. This was done to produce the solid curve in
\fig~\ref{warnfig}. One might be tempted to try and determine the \fwhm\ and
the background simultaneously in the fitting procedure. However, this gives
very bad results as is indicated by the dotted curve in \fig~\ref{warnfig}.
Here a gaussian $a_1 \exp( -a_2 r^2 ) + a_3$ was fitted to the constant
surface brightness disk, treating $a_3$ as a free parameter. One can readily
see that the results are poor. The radius of the fitting region has to be very
large to get even moderately accurate results. The cause of this is that the
fit is not punished for the fact that the gaussian drops below the background
at infinity, which in turn gives it the freedom to make the gaussian wider
and thus obtain a better fit inside the integration region.

For very small values of $\beta$, SB find higher values for the conversion
factor than this study. This is probably an artifact of the fitting function
adopted by SB. It was found to give lower quality fits than the fitting
function adopted in this paper, while it has the same number of free
parameters.
The conversion factor
$\gamma(\beta)$ has a first derivative which is zero at $\beta=0$, as is also
the case for the fitting function adopted in this paper. This is however not
the case for the fitting function adopted by SB.
Since it is
impossible to compute the conversion factor for very small $\beta$ using
method~B (as was done by SB), their fit will not be constrained for those
values and thus
their choice of fitting function will give results
which are too high near $\beta$ = 0.

When the results of PW are viewed for large beam sizes, it can be seen that
the conversion factor is roughly 1.7 in the disk case, and roughly 1.9 in the
sphere case. This had led to the popular notion that the deconvolved \fwhm\
should be multiplied by 1.8 to obtain the correct diameter, independent of
beam size and intrinsic surface brightness distribution. The popularity of
this assumption should in all probability be attributed to its simplicity.
However, from the previous discussion it becomes clear that quite substantial
errors can be made this way, as was already pointed out by PW. The magnitude
of this error can easily be greater than the observational uncertainties in
the measurement itself. Hence this simplification cannot be justified. This
implies that assumptions about the intrinsic surface brightness profile are
unavoidable. Since there obviously is room for discussion about these
assumptions, the author would like to urge all observers to publish besides
the derived angular diameter (which depends on these assumptions) also both the
deconvolved \fwhm\ (which does not depend on these assumptions) {\em and} the
beam size.

\section{Optical depth effects}
\label{op:depth}

In the previous sections several simple geometries have been studied where an
analytic expression for the intrinsic surface brightness profile can be
assumed. However, in the following sections more realistic geometries will be
studied. These will be based on a photo-ionization model of a planetary
nebula. The procedure to calculate the conversion factors in this case is as
follows. A table with the emissivity and the absorption coefficient as a
function of distance to the star is calculated with a modified version of the
photo-ionization code \cld\ 84.12a (Ferland 1993). When the emissivities and
the absorption coefficients are known, the radiative transport equation can be
integrated numerically, assuming spherical symmetry and neglecting scattering
processes. This is done by a separate code which yields an intrinsic surface
brightness profile. This profile can then be used to determine the conversion
factors using the procedure already described in \sct~\ref{comput:gauss}.

\begin{table}
\caption{The parameters for calculating conversion factors in different
observing modes. The conversion factors for gaussian deconvolution can
be calculated using \req{fitfunc}, the conversion factors for second moment
deconvolution can be calculated using \req{gam2use}.}
\label{observtab}
\begin{tabular}{lllll}
\hline
case & \sctr{$a_{1}$} & \sctr{$a_{2}$} & \sctr{$a_{3}$} & \sctr{$\tau\m$} \\
\hline
radio 2~cm   &   0.3428 &   0.7853 &   1.6222 &  1.44($-$2) \\
radio 6~cm   &   0.3429 &   0.7860 &   1.6067 &  1.46($-$1) \\
radio 20~cm  &   0.3462 &   0.7892 &   1.4767 &  1.81($+$0) \\
radio 60~cm  &   0.3529 &   0.7870 &   1.3708 &  1.79($+$1) \\
radio 200~cm &   0.3433 &   0.7854 &   1.3753 &  2.18($+$2) \\
\\
H$\alpha$            &   0.3605 &   0.7822 &   1.4937 &  4.53($-$1) \\
H$\alpha$+\fb{N}{ii} &   0.3569 &   0.7867 &   1.3656 &  4.53($-$1) \\
H$\beta$             &   0.3644 &   0.7817 &   1.4842 &  6.30($-$1) \\
\fb{O}{iii}          &   0.4083 &   0.7892 &   1.3176 &  6.11($-$1) \\
\hline
\end{tabular}
\par\nt{\m\x{4}The optical depth is measured from the centre of the nebula to
the outer edge.}
\end{table}

\begin{figure}
\begin{center}
\mbox{\epsfxsize=0.45\textwidth\epsfbox[21 317  534 668]{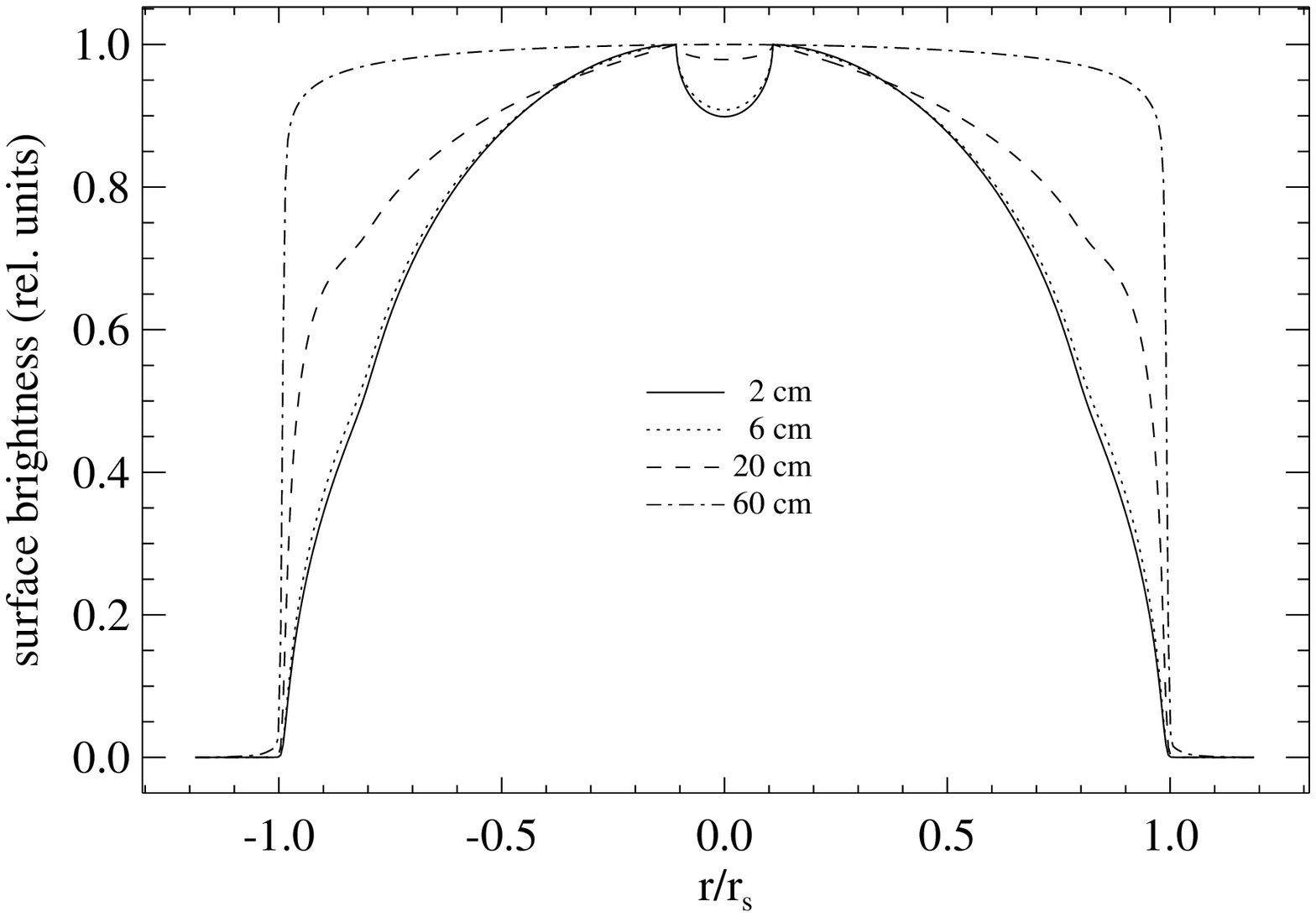}}
\vskip\floatsep
\mbox{\epsfxsize=0.45\textwidth\epsfbox[21 317  534 668]{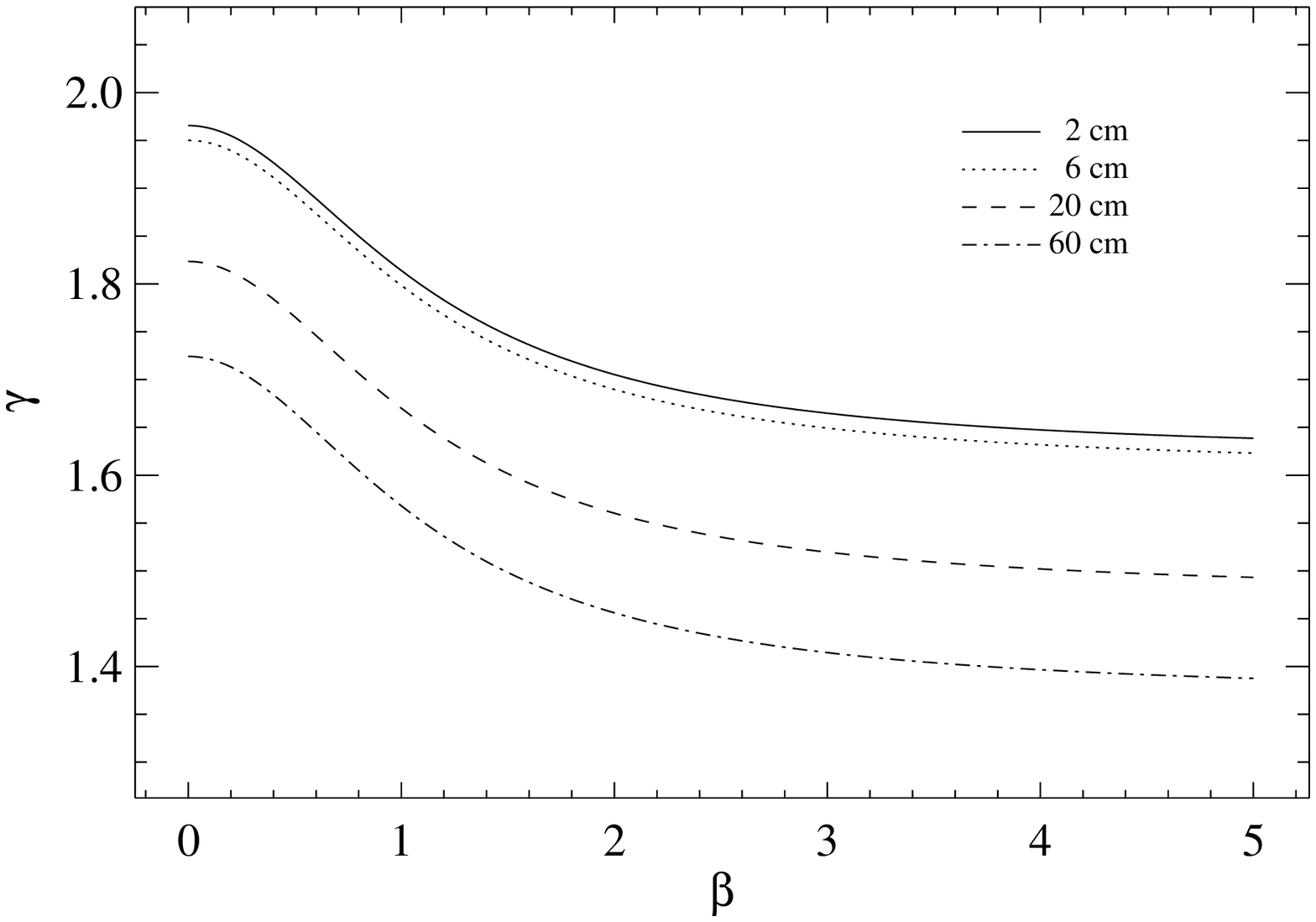}}
\caption{The surface brightness profiles (upper panel) and the conversion
factors (lower panel) computed at wavelengths of 2~cm, 6~cm, 20~cm and 60~cm.
The maximum surface brightness has been normalized to unity.}
\label{shape1}
\end{center}
\end{figure}

To study the effect of varying optical depth with wavelength, the results of
the planetary nebula model at various wavelengths in the radio regime have
been used. The physical process responsible for emission and absorption at
these wavelengths is free-free interaction between protons and electrons. It
is a well known fact that the optical depth $\tau$ due to free-free absorption
increases towards longer wavelengths as $\tau \propto \lambda^{2.1}$ (e.g.,
Pottasch 1984). The input parameters for \cld\ were taken from the
\astrobj{NGC~7027} model discussed in Beintema et al. \cite{c3:beintema}. This
nebula was chosen solely for the purpose of getting a realistic model. The
fact that \astrobj{NGC~7027} is quite large and thus well resolved in most
observations is of no importance. One could assume that another nebula is
modelled which is very similar to \astrobj{NGC~7027}, but at a much larger
distance. Using this model, the shape of the surface brightness profile at
various wavelengths could be determined. It turns out that the surface
brightness profile is very sensitive to optical depth effects. To illustrate
this, various profiles are shown in the upper panel of \fig~\ref{shape1}. The
conversion factors to obtain the true angular diameter at the various
wavelengths are shown in the lower panel of \fig~\ref{shape1}. The nebula is
only marginally optically thick at 6~cm, but this already has a noticeable
effect on the conversion factor. At 20~cm the nebula is mildly optically
thick, and this has a strong effect on the conversion factor.

The effect may be less in other nebulae.
However, in general this cannot be assumed {\it a priori}, and thus
care should be taken when comparing measurements taken at different
wavelengths. This is not only true because of the effects described above, but
also because at different wavelengths the beam size will be different, hence
$\beta$ will be different and this will influence the conversion factor as
well when gaussian deconvolution is used.

\section{Imaging in emission lines}
\label{hydrogen}

\begin{figure}
\begin{center}
\mbox{\epsfxsize=0.45\textwidth\epsfbox[21 317  534 668]{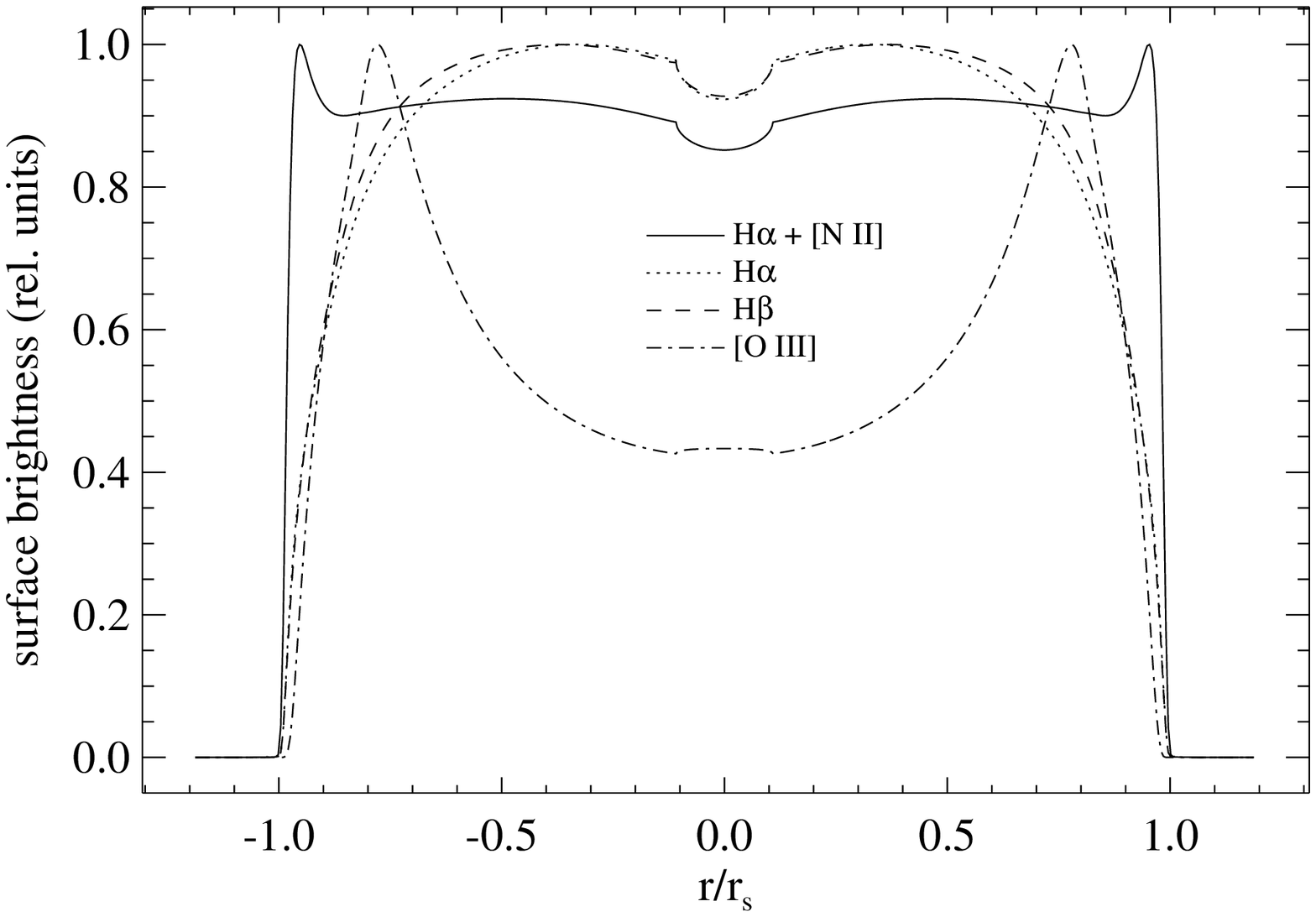}}
\vskip\floatsep
\mbox{\epsfxsize=0.45\textwidth\epsfbox[21 317  534 668]{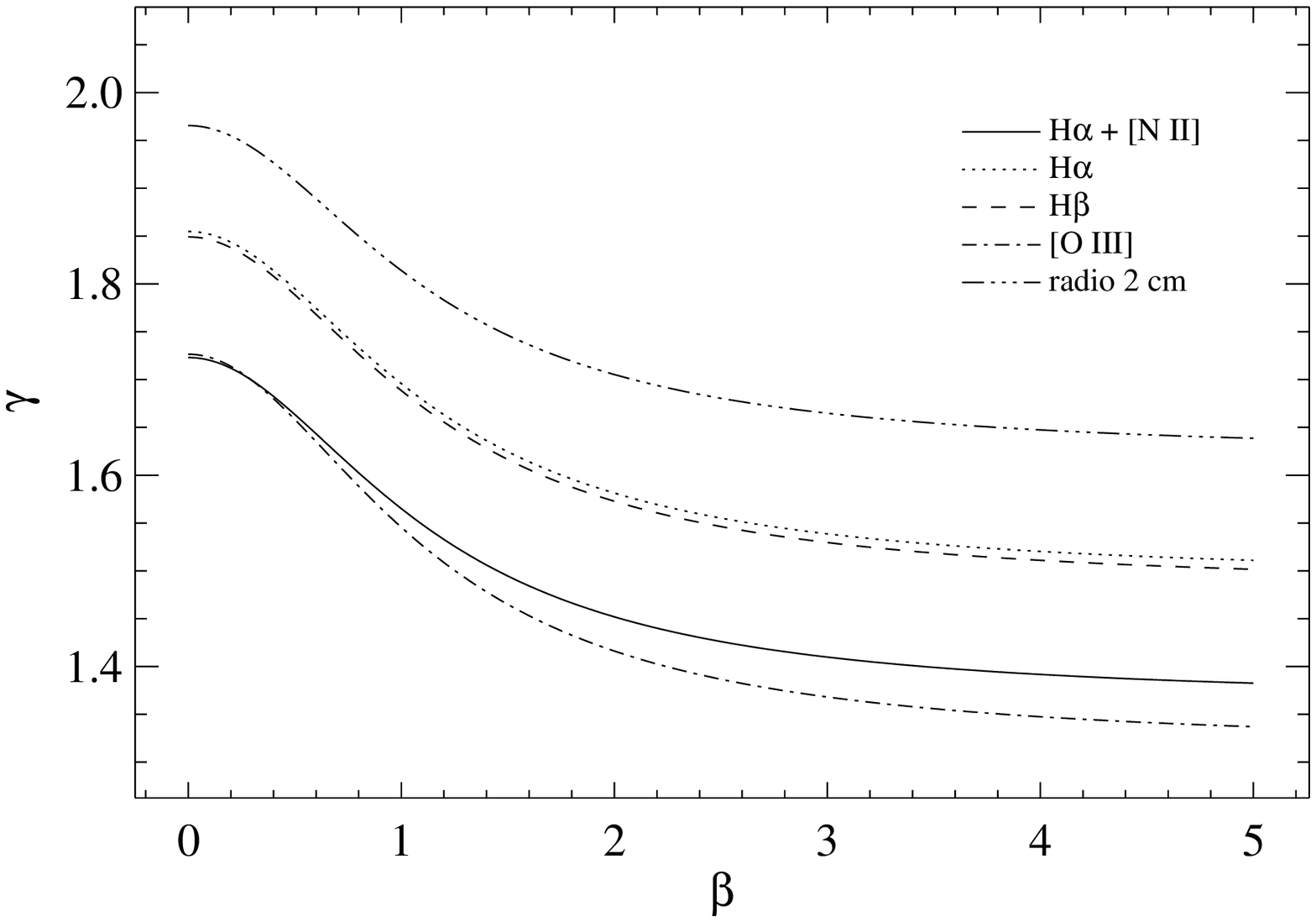}}
\caption{The intrinsic surface brightness profiles (upper panel) and the
conversion factors (lower panel) in various emission lines. The maximum
surface brightness has been normalized to unity.
The conversion factors for the 2~cm radio image are also shown for
reference.}
\label{shape2}
\end{center}
\end{figure}

Since optical images are also used to determine the radius of a nebula, the
conversion factors for these images will be investigated as well. For this the
same model was used as discussed in the previous section. The emissivity was
assumed to be the volume emissivity of the following emission lines: \ha, \hb,
\fb{N}{ii} \w 6548, \w 6584 and \fb{O}{iii} \w 5007. The contribution of
continuum emission in the images was neglected. When solving the radiative
transport equations, continuum optical depth effects were included, but not
line optical depth effects. Next, the surface brightness profiles were
computed for images taken in pure \ha\ light, in \ha+\fb{N}{ii} \w\w 6548,
6584, in \hb\ and in \fb{O}{iii} \w 5007. The results are shown in the upper
panel of \fig~\ref{shape2}. Finally, the conversion factors were computed for
these profiles. The parameters for the fits are given in \tbl~\ref{observtab},
the curves are shown in the lower panel of \fig~\ref{shape2}.

The first thing that can be noticed is that the surface brightness profiles in
the various emission lines look completely different and that they also differ
from the radio surface brightness profiles shown in \fig~\ref{shape1}. Also
the positions for the peak surface brightness are completely different. This
implies that even in well-resolved images, the measured diameter can be
different depending on which line is used. This is a well known effect
caused by ionization stratification.
Given this, it can already be expected that the conversion factors
should be different for the various images as well, which is indeed confirmed.
The conversion factors for images in pure \ha\ and \hb\ light are nearly
identical, which can be expected from the fact that the relative level
populations depend only mildly on electron temperature and density. However,
this situation completely changes when the \fb{N}{ii} lines are included in
the passband of the \ha\ filter (as is usually the case). This has a
considerable effect on the conversion factor. Also the conversion factors for
the \fb{O}{iii} image are quite different from the \hb\ case, but, by chance,
nearly coincide with the values for the \ha+\fb{N}{ii} image. It can be seen
that all conversion factors for the optical images are considerably smaller
than the values for the (optically thin) 2~cm radio image.

From this the conclusion can be drawn that angular diameters measured in
different images (including radio observations) can {\em not} be compared
directly, even when they are well-resolved. In general the conversion factors
for the various images will be different, depending on the intrinsic surface
brightness profile and the beam size of the observation. For the cases that
have been studied above, differences exceeding 30~\% are possible.

\section{Non-constant density geometries}
\label{non:constant}

\begin{figure}
\begin{center}
\mbox{\epsfxsize=0.45\textwidth\epsfbox[21 317  534 668]{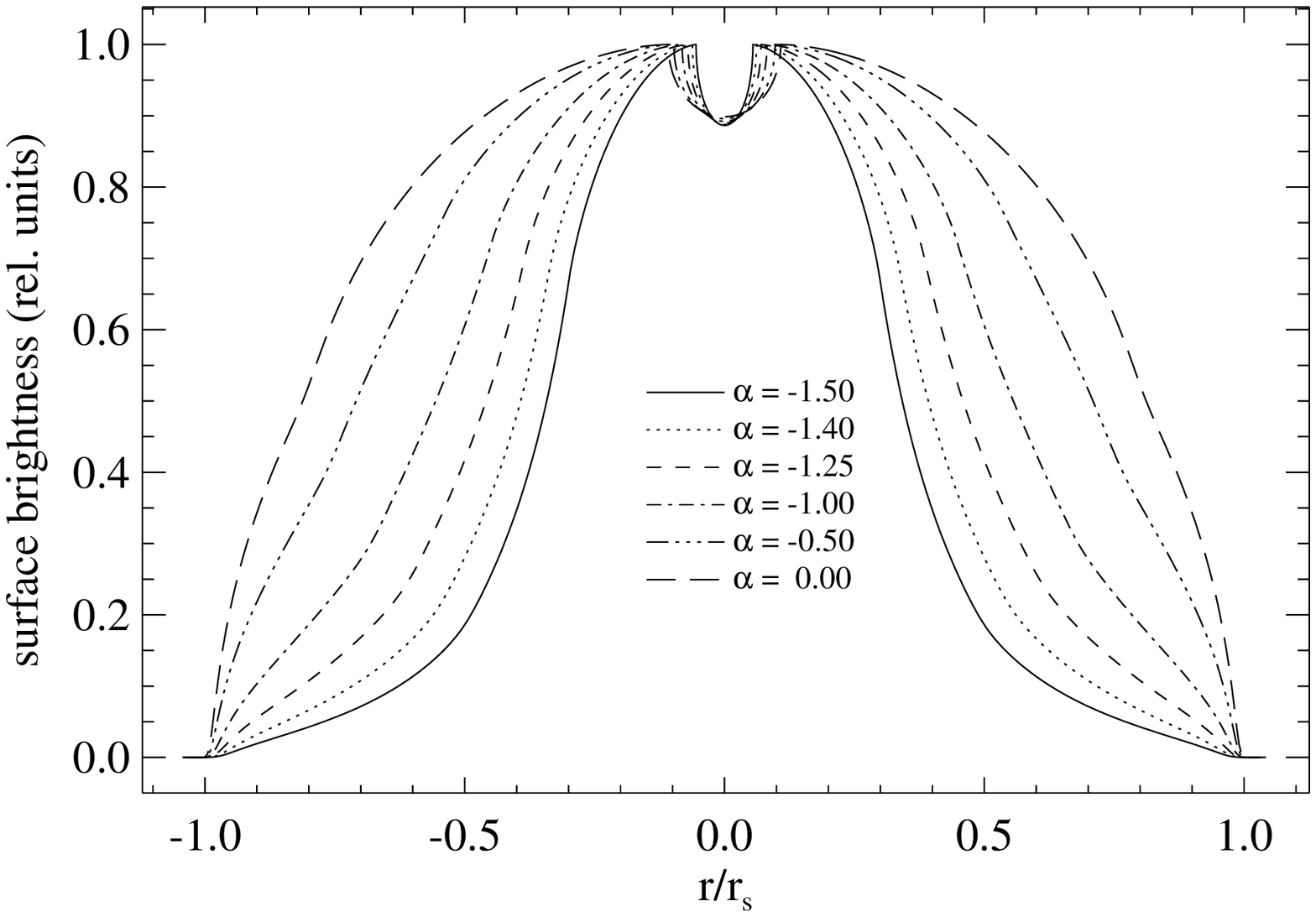}}
\vskip\floatsep
\mbox{\epsfxsize=0.45\textwidth\epsfbox[21 317  534 668]{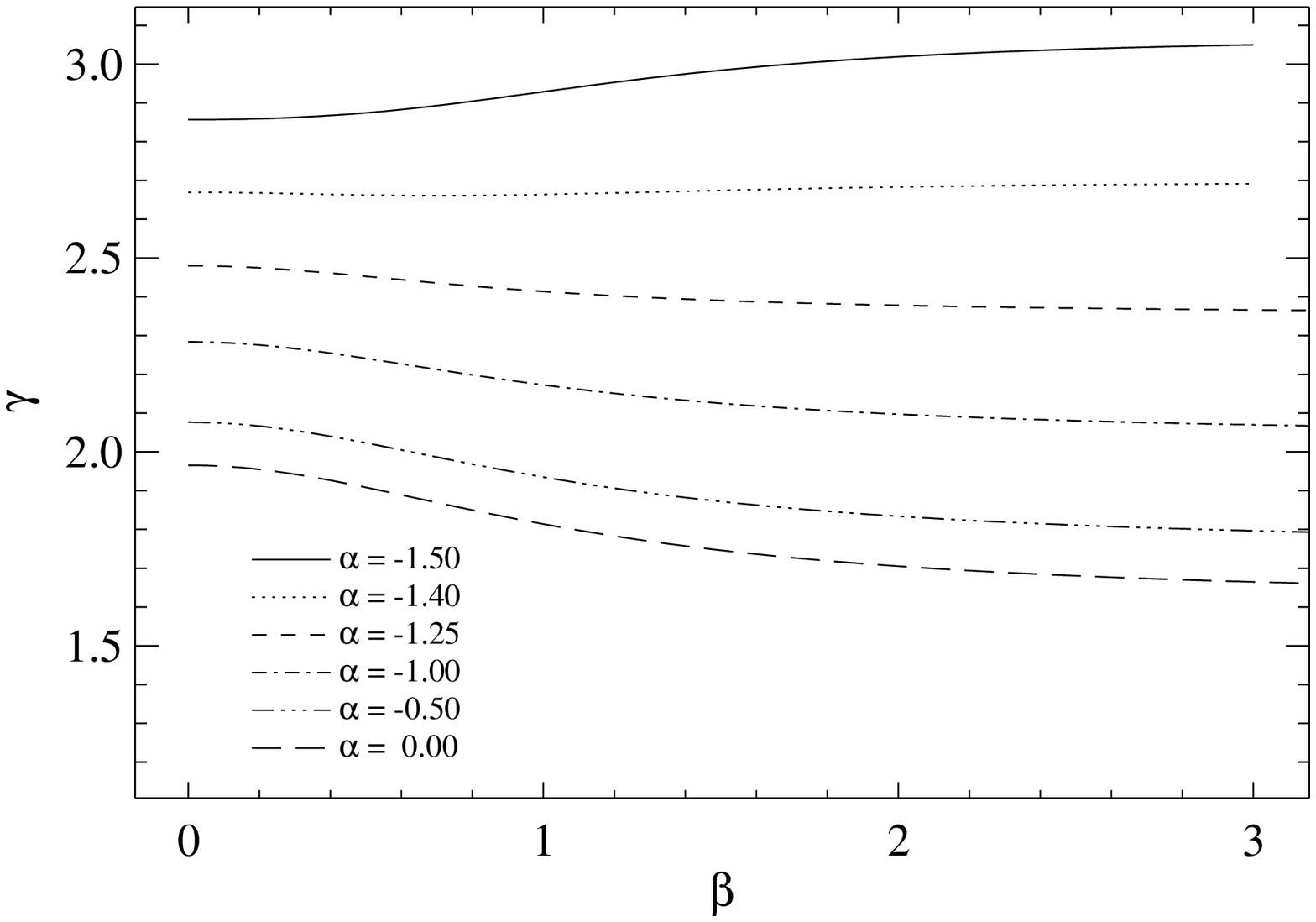}}
\caption{The intrinsic surface brightness profiles (upper panel) and the
conversion factors (lower panel) for different density distributions. The
maximum surface brightness has been normalized to unity.}
\label{shape3}
\end{center}
\end{figure}

\begin{table}
\caption{The parameters for calculating the conversion factors for various
density profiles. The conversion factors for gaussian deconvolution can
be calculated using \req{fitfunc}, the conversion factors for second moment
deconvolution can be calculated using \req{gam2use}. The diameter is given in
arbitrary units.}
\label{alpha:tab}
\begin{tabular}{rlrrl}
\hline
\sctr{$\alpha$} & \sctr{$\dia$} & \sctr{$a_1$} & \sctr{$a_2$} & \sctr{$a_3$} \\
\hline
$0.00$  & 1.137 &   0.3428  &   0.7853 &   1.6222 \\
$-0.50$ & 1.261 &   0.3188  &   0.7911 &   1.7573 \\
$-1.00$ & 1.504 &   0.2418  &   0.8397 &   2.0415 \\
$-1.25$ & 1.747 &   0.1253  &   1.1091 &   2.3545 \\
$-1.40$ & 1.994 &$-$0.078\f &   0.078\f&   2.740  \\
$-1.50$ & 2.250 &$-$0.252\f &   0.476\f&   3.103  \\
\hline
\end{tabular}
\end{table}

The nebular model that has been discussed so far assumes a constant hydrogen
density within the ionized region. This automatically leads to a well-defined
boundary of the nebula. To investigate how the conversion factors behave for
nebulae with soft boundaries, the density law was changed in the following
way. For the inner parts of the nebula [lg($r$/cm) $<$ 16.6] the density law
remained unchanged (i.e.\ constant), but for the outer parts an $r^\alpha$ law
was assumed. This roughly models a nebula where the inner parts have undergone
hydrodynamic interactions with the fast post-AGB wind, but where the outer
parts are not yet disturbed. Values of $\alpha$ were used ranging from 0.0
down to $-1.5$. All the other parameters of the model were kept unaltered. For
all models a radio image was calculated at a wavelength of 2~cm. The surface
brightness profiles and conversion factors are shown in \fig~\ref{shape3}. In
\tbl~\ref{alpha:tab} the parameters for the fits to the conversion factors
together with the true angular diameters of the nebulae can be found.

The first thing that can be noticed is that the conversion factor becomes
increasingly larger as $\alpha$ becomes more negative. This can be easily
understood when the nature of the changes to the density profile are viewed.
Since a large part of the constant density regime within the ionized region
remains unaltered when $\alpha$ is changed, also the high surface brightness
core of the nebula remains more or less unaltered. Since the \fwhm\ is mainly
determined by the core region, the deconvolved \fwhm\ will not change much as
a function of $\alpha$. However, the \str\ radius is very sensitive to
$\alpha$, as can be seen in \tbl~\ref{alpha:tab}. This implies that the
conversion factor must also be a sensitive function of $\alpha$. If the \str\
radius is chosen as a physically meaningful radius, this makes it almost
impossible to measure this radius from low resolution observations if density
distributions similar to the ones discussed here are suspected. It is the
authors opinion that in such a case the only meaningful thing to do is to
publish the deconvolved \fwhm\ and the beam size and not to make any attempt
to calculate the \str\ radius.

\section{A new algorithm to determine the intrinsic FWHM}
\label{new:method}

\begin{table*}
\caption{The results from the implicit deconvolution method for a disk, sphere
and shell geometry as defined in \sct~\ref{simple}. The number following the
name of the geometry indicates the photon count at the maximum surface
brightness. $\Phb$ indicates the \fwhm\ of the beam, $\Phi$ the \fwhm\ of the
convolved profile and $\Phi_{\rm in}$ the intrinsic \fwhm. In each case the
measurements were repeated 24 times using a different seed for the random
generator; the quoted \fwhm\ is the average of all cases where the method
converged, $\sigma$ indicates the standard deviation (68~\% confidence
interval) of an individual measurement. Cases where the implicit deconvolution
method converged less than 12 times are omitted from the table. In all cases
the beam size is assumed to be measured with high accuracy.}
\label{intrin}
\scriptsize
\begin{tabular}{r@{\x{6}}r@{\x{2}}r@{\x{2}}r@{\x{2}}r@{\x{0}}r@{\x{6}}r@{\x{2}}r@{\x{2}}r@{\x{2}}r@{\x{0}}r@{\x{6}}r@{\x{2}}r@{\x{2}}r@{\x{2}}r@{\x{0}}r}
\hline
& \multicolumn{4}{c}{disk -- 500} & & \multicolumn{4}{c}{disk -- 5000} & & \multicolumn{4}{c}{disk -- 50000} \\
\ctr{1}{$\Phb$}{5} & \sctr{$\Phi$} & \sctr{$\sigma$} & \sctr{$\Phi_{\rm in}$} & \sctr{$\sigma$} & & \sctr{$\Phi$} & \sctr{$\sigma$} & \sctr{$\Phi_{\rm in}$} & \sctr{$\sigma$} & & \sctr{$\Phi$} & \sctr{$\sigma$} & \sctr{$\Phi_{\rm in}$} & \sctr{$\sigma$} \\
\cline{2-5} \cline{7-10} \cline{12-15}
0.25 &  1.493  &  0.006 &  1.486  &  0.006 & &  1.4930  &  0.0014 &  1.4858  &  0.0014 & &  1.4934  &  0.0005 &  1.4861  &  0.0006 \\
0.35 &  1.501  &  0.011 &  1.486  &  0.011 & &  1.5021  &  0.0025 &  1.4866  &  0.0026 & &  1.5021  &  0.0008 &  1.4867  &  0.0009 \\
0.50 &  1.520  &  0.012 &  1.485  &  0.013 & &  1.5200  &  0.0038 &  1.4856  &  0.0041 & &  1.5211  &  0.0013 &  1.4868  &  0.0013 \\
0.71 &  1.563  &  0.018 &  1.481  &  0.023 & &  1.5672  &  0.0054 &  1.4871  &  0.0068 & &  1.5686  &  0.0016 &  1.4887  &  0.0020 \\
1.00 &  1.673  &  0.032 &  1.476  &  0.043 & &  1.6808  &  0.0084 &  1.4902  &  0.0132 & &  1.6822  &  0.0026 &  1.4921  &  0.0039 \\
1.41 &  1.917  &  0.045 &         &        & &  1.9261  &  0.0116 &          &         & &  1.9276  &  0.0035 &          &         \\
\hline
& \multicolumn{4}{c}{sphere -- 500} & & \multicolumn{4}{c}{sphere -- 5000} & & \multicolumn{4}{c}{sphere -- 50000} \\
\ctr{1}{$\Phb$}{5} & \sctr{$\Phi$} & \sctr{$\sigma$} & \sctr{$\Phi_{\rm in}$} & \sctr{$\sigma$} & & \sctr{$\Phi$} & \sctr{$\sigma$} & \sctr{$\Phi_{\rm in}$} & \sctr{$\sigma$} & & \sctr{$\Phi$} & \sctr{$\sigma$} & \sctr{$\Phi_{\rm in}$} & \sctr{$\sigma$} \\
\cline{2-5} \cline{7-10} \cline{12-15}
0.25 &  1.296  &  0.008 &  1.284  &  0.008 & &  1.2946  &  0.0019 &  1.2832  &  0.0019 & &  1.2949  &  0.0007 &  1.2834  &  0.0011 \\
0.35 &  1.313  &  0.012 &  1.288  &  0.013 & &  1.3132  &  0.0034 &  1.2885  &  0.0037 & &  1.3137  &  0.0011 &  1.2890  &  0.0012 \\
0.50 &  1.335  &  0.012 &  1.282  &  0.014 & &  1.3365  &  0.0042 &  1.2842  &  0.0048 & &  1.3379  &  0.0013 &  1.2857  &  0.0015 \\
0.71 &  1.395  &  0.022 &  1.277  &  0.031 & &  1.4013  &  0.0058 &  1.2862  &  0.0085 & &  1.4025  &  0.0018 &  1.2875  &  0.0026 \\
1.00 &  1.536  &  0.033 &         &        & &  1.5433  &  0.0087 &  1.2956  &  0.0169 & &  1.5447  &  0.0027 &  1.2926  &  0.0054 \\
1.41 &  1.815  &  0.043 &         &        & &  1.8240  &  0.0114 &          &         & &  1.8254  &  0.0035 &          &         \\
\hline
& \multicolumn{4}{c}{shell 0.8 -- 500} & & \multicolumn{4}{c}{shell 0.8 -- 5000} & & \multicolumn{4}{c}{shell 0.8 -- 50000} \\
\ctr{1}{$\Phb$}{5} & \sctr{$\Phi$} & \sctr{$\sigma$} & \sctr{$\Phi_{\rm in}$} & \sctr{$\sigma$} & & \sctr{$\Phi$} & \sctr{$\sigma$} & \sctr{$\Phi_{\rm in}$} & \sctr{$\sigma$} & & \sctr{$\Phi$} & \sctr{$\sigma$} & \sctr{$\Phi_{\rm in}$} & \sctr{$\sigma$} \\
\cline{2-5} \cline{7-10} \cline{12-15}
0.25 &  1.637  &  0.007 &  1.633  &  0.007 & &  1.6366  &  0.0014 &  1.6326  &  0.0014 & &  1.6370  &  0.0004 &  1.6330  &  0.0004 \\
0.35 &  1.641  &  0.010 &  1.632  &  0.010 & &  1.6415  &  0.0023 &  1.6326  &  0.0024 & &  1.6416  &  0.0007 &  1.6328  &  0.0008 \\
0.50 &  1.652  &  0.012 &  1.631  &  0.012 & &  1.6522  &  0.0037 &  1.6311  &  0.0038 & &  1.6532  &  0.0012 &  1.6322  &  0.0012 \\
0.71 &  1.682  &  0.014 &  1.628  &  0.016 & &  1.6845  &  0.0045 &  1.6313  &  0.0052 & &  1.6858  &  0.0014 &  1.6327  &  0.0016 \\
1.00 &  1.766  &  0.029 &  1.625  &  0.037 & &  1.7747  &  0.0077 &  1.6336  &  0.0095 & &  1.7760  &  0.0023 &  1.6347  &  0.0033 \\
1.41 &  1.981  &  0.044 &         &        & &  1.9917  &  0.0114 &          &         & &  1.9932  &  0.0035 &          &         \\
\hline
\end{tabular}
\end{table*}

In Paper~I it has been shown that the deconvolved \fwhm\ obtained from second
moment deconvolution is independent of the beam size, and therefore equal to
the (second moment) \fwhm\ of the intrinsic profile. Hence this method has the
advantage that the \fwhm\ of the unconvolved profile can be obtained
directly from the observations
without making any assumptions about the shape of the profile.
Both from Paper~I and this
paper it has become clear that gaussian deconvolution does {\em not} have this
property and assumptions about the intrinsic profile are always necessary.
From the previous section it is also clear that such assumptions are not
always warranted. This situation is not very satisfactory, and therefore a new
algorithm will be presented here which will remedy this problem. This
algorithm will yield the (gaussian fit) \fwhm\ of the intrinsic profile, given
the observed profile and the \fwhm\ of the beam, without making any
assumptions about the intrinsic profile. To use this method one has to
determine the radial moments $c_{2n}\p$ of the observed (i.e., convolved)
profile and convert them to the radial moments of the intrinsic (i.e.,
unconvolved) profile $c_{2n}$.
The radial moments $c_{2n}\p$ are related to $c_{2n}$ by \req{impl:expl}.
\begin{equation}
   c_{2n}\p = \sum_{k=0}^n \frac{n!^2}{(n-k)!\,k!^2} \, \frac{c_{2k}}{p^{n-k}};
   \x{3} p \equiv \frac{4\ln2}{\Phb^2} \x{3} \Rightarrow
\label{impl:expl}
\end{equation}
\[
   c_0\p = c_0, \x{3} c_2\p = c_2 + \frac{c_0}{p}, \x{3}
   c_4\p = c_4 + \frac{4c_2}{p} + \frac{2c_0}{p^2}, \x{3} \cdots
\]
This relation can easily be inverted
\[
   c_0 = c_0\p, \x{3} c_2 = c_2\p - \frac{c_0\p}{p}, \x{3}
   c_4 = c_4\p - \frac{4c_2\p}{p} + \frac{2c_0\p}{p^2}, \x{3} \cdots
\]
These relations constitute the actual deconvolution. In order to obtain the
\fwhm\ of the deconvolved profile (which we will call the intrinsic \fwhm),
the values of $c_{2n}$ resulting from this calculation have to be substituted
in \req{gen:inf}. A derivation of this algorithm is presented in \scts~\alter\
and \secmom\ of Paper~I. This method uses the fact that information about the
intrinsic profile is implicitly contained in the radial moments of the
convolved profile, and will therefore be called implicit deconvolution.

In the remainder of the section this method will be tested on artificial data.
To this end the three simple geometries that where already discussed in
\sct~\ref{previous} will be used.
In order to simulate realistic observing conditions,
these surface brightness distributions were
convolved with a gaussian beam of prescribed width and observational noise 
was added using a Poisson noise generator. For this the
photon count of the peak surface brightness was prescribed. Additionally a
read-out noise of 6 counts was assumed. The pixel scale of the CCD was chosen
such that in each case the \fwhm\ of the beam corresponded to approximately 5
pixels. The results of the tests are shown in \tbl~\ref{intrin}. Before the
results of the implicit deconvolution method will be discussed, first two
remarks will be made concerning the alternative method to measure the \fwhm\
(given by \req{gen:inf}).

In Paper~I it was shown that this method yields identical results to a
gaussian fit algorithm. This is true when the profile is perfectly sampled.
However, in real data the profile is only partially sampled and this may
lead to small discrepancies between the two methods on the order of the
measurement uncertainty in the \fwhm.

A second concern is that, due to noise,
negative values for pixels can occur in the low surface brightness areas.
The theory presented in Paper~I is strictly speaking not valid for such
profiles. In practice however, this was found not to give any problems. All
measurements of the convolved \fwhm\ ($\Phi$) presented in \tbl~\ref{intrin}
converged, even in the poorest signal-to-noise conditions.

The results in \tbl~\ref{intrin} show that the implicit deconvolution method
gives stable results in realistic conditions, provided the beam size is less
than roughly $2/3$ of the observed \fwhm. The accuracy of the resulting
intrinsic \fwhm\ can be judged by comparing it to the results in
\tbl~\ref{tab:inf}. By looking at the high signal-to-noise results, one can
see that small discrepancies can occur. These can be attributed to the fact
the observed profile was only sampled with a very small number of pixels, as was
discussed above. It is interesting to note that the implicit deconvolution
method is not hampered by low signal-to-noise conditions, in the sense that
the maximum beam size for which the method still works is hardly affected by
low signal-to-noise conditions. It is therefore concluded that the implicit
deconvolution method works well in realistic conditions, even when the
signal-to-noise is low, provided that the beam size is less than roughly $2/3$
of the observed \fwhm\ and the beam profile can be approximated by a gaussian.
It can be a good alternative for gaussian deconvolution of partially resolved
sources, since it requires no assumptions on the intrinsic surface brightness
distribution. However, in order to convert the intrinsic \fwhm\ of the source
into the \str\ diameter, knowledge about the intrinsic surface brightness
distribution is still needed. A {\sc fortran} program implementing the
algorithm discussed in this section is available from
\url{ftp://gradj.pa.uky.edu/pub/peter/genfit.f}.

\section{Conclusions}
\label{concl:iv}

In this work conversion factors have been determined to convert the
deconvolved \fwhm\ of a partially resolved nebula to its true diameter. It was
already found by Mezger \& Henderson \cite{c3:mezger} that this conversion
factor depends on the (assumed) intrinsic surface brightness profile of the
nebula. In a subsequent study by Panagia \& Walmsley \cite{c3:pana} it was
found that the conversion factor also depends on the beam size of the
observation when gaussian deconvolution is used. This paper expands on
previous work in that an alternative method for deconvolving the \fwhm, second
moment deconvolution, is studied for the first time. Also the influence of the
intrinsic surface brightness profile on the conversion factor is studied in
more detail. The following recommendations and conclusions were reached.
Unless explicitly noted otherwise, they are valid both for gaussian and second
moment deconvolution and also for observations at arbitrary wavelengths.
\begin{enumerate}
\item
When making a gaussian fit to a surface brightness profile, it is recommended
to use a diameter for the fitting region which is at least 3 times the \fwhm\
diameter of the fit in all cases.
A larger fitting region should be used if
extended faint emission is present.
In order to obtain an accurate value for the \fwhm\ of an observed source,
it is essential that the global background in the image is subtracted
first so that it can assumed to be zero in the fitting procedure.
Trying to determine the \fwhm\ and the background simultaneously
in the fitting procedure is found to give very poor results.
\item
The deconvolved \fwhm\ derived using gaussian fits is in general not equal to
the deconvolved \fwhm\ derived using second moments. Hence the conversion
factors will also be different in both cases. For second moment deconvolution,
the conversion factor is {\em independent} of the beam size. Its value is in
all cases equal to the conversion factor for the gaussian deconvolution method
in the limit for infinitely large beams. The conversion factor for second
moment deconvolution does depend on the assumed surface brightness profile.
\item
The conversion factor is very sensitive to the adopted intrinsic surface
brightness profile. Differences up to 40~\% can be found for constant
emissivity shells with different inner radii. Hence, great care should be
taken when making a choice for the intrinsic surface brightness distribution.
\item
Because of this, observers are urged to publish, besides the \str\ diameter of
the nebula, the deconvolved \fwhm, the method used (i.e.\ gaussian or second
moment deconvolution) {\em and} the beam size.
\item
The conversion factor is very sensitive to optical depth effects, so care
should be taken when comparing observations made at different wavelengths.
This is especially the case for radio observations. Differences of several
tens of percent are possible.
\item
For optical observations the conversion factor depends on which emission line
is chosen. This is partly due to ionization stratification and this results in the
fact that even well-resolved images in different emission lines can yield
different diameters. It is also caused by the fact that the intrinsic surface
brightness profile is different in different emission lines. Again differences
of several tens of percent are possible. Hence care should be taken when
comparing optical and radio measurements.
\item
Nebulae which have a power law drop-off in their density distribution usually
do not have a well-defined outer edge and the \str\ radius will be situated in
the faint surface brightness regions of the nebula. For such nebulae the
conversion factor can become very large and is very sensitive to the assumed
intrinsic surface brightness distribution of the nebula. Since this
distribution can in general not be assessed accurately, it is not meaningful
to apply a conversion factor and only the deconvolved \fwhm\ and the beam size
should be published.
\end{enumerate}

\noindent
Finally, in this paper a new algorithm has been presented which allows the
determination of the intrinsic \fwhm\ of the source, using only the observed
surface brightness distribution and the \fwhm\ of the beam. More in
particular, no assumptions with regard to the intrinsic surface brightness
distribution are needed. This makes the method a good alternative for gaussian
deconvolution. Tests show that the implicit deconvolution method works well in
realistic conditions, even when the signal-to-noise is low, provided that the
beam size is less than roughly $2/3$ of the observed \fwhm\ and the beam
profile can be approximated by a gaussian.

\section*{Acknowledgments}
The author would like to thank G.C. Van de Steene for inspiring this research.
G.C. Van de Steene, K. \mbox{Kuijken} and the referee A.A. \zijlstra\ are thanked for
critically reading the manuscript. The photo-ionization code \cld\ has been
used, written by Gary Ferland and obtained from the University of Kentucky,
USA. The author was supported by NFRA grant 782--372--033 during his stay in
Groningen, and by the NSF through grant no.\ AST 96--17083 during his stay
in Lexington.

\end{document}